\documentclass[letter,aps,prb,twocolumn]{revtex4}
\usepackage[latin1]{inputenc}
\usepackage{amsmath}
\usepackage{txfonts}
\usepackage{bm}
\usepackage{graphicx}
\newcommand{\ie}{i.~e.,} 
\newcommand{\eg}{e.~g.,}
\newcommand{\hc}{{\text{H.\ c.}}}  
\newcommand{\ket}[1]{|#1\rangle}

\newcommand{\matel}[3]{\langle#1|\,#2\,|#3\rangle} 

\newcommand{\gc}{{\mathcal G}_2} 
\newcommand{\gset}{G_{}^S}

\newcommand{\cd}{c_{d}^{\phantom{\dagger}}} 
\newcommand{\fzero}{f_{0}^{\phantom{\dagger}}} 
\newcommand{\cdd}{c_{d}^\dagger}

\newcommand{\ed}{\varepsilon_d}

\newcommand{\lp}{\bm{\left(}} 
\newcommand{\rp}{\bm{\right)}}
\newcommand{\ha}{H_A}
\newcommand{\hs}{H_A^S}

\newcommand{\mc}[1]{\mathcal{#1}}
\newcommand{\vw}{W}

\newcommand{\za}{\mathcal{Z}}

\newcommand{\gammaw}{\Gamma_{w}}
\newcommand{\deltaw}{\delta_w}

\newcommand{\te}[1]{$\times10^{-#1}$}
\newcommand{\ti}{t^{\phantom{A}}}

\begin{document}
\title{Universal zero-bias conductance for the single electron
  transistor. II: Comparison with numerical results}
\author{M. Yoshida} \affiliation{Departamento de F\'{\i}sica,
  Instituto de Geoci\^{e}ncias
  e Ci\^encias Exatas,\\ Universidade Estadual Paulista, 13500, Rio
  Claro, SP, Brazil}
\author{A. C. Seridonio}
\altaffiliation{Present address: Instituto de Física\\ Universidade Federal
  Fluminense, Niter\'oi, 24210-346, RJ- Brazil}

\author{L.~N.~Oliveira} \affiliation{Departamento de F\'{\i}sica e
  Inform\'{a}tica, Instituto
  de F\'{\i}sica de S\~{a}o Carlos, \\ Universidade de S\~{a}o Paulo,
  369, S\~{a}o Carlos, SP, Brazil}

\begin{abstract}
  A numerical renormalization-group survey of the zero-bias electrical
  conductance through a quantum dot embedded in the conduction path of
  a nanodevice is reported. The results are examined in the light of
  a recently derived linear mapping between the temperature-dependent
  conductance and the universal function describing the conductance
  for the symmetric Anderson model. A gate potential applied to the
  conduction electrons is known to change markedly the transport
  properties of a quantum dot side-coupled to the conduction path; in
  the embedded geometry here discussed, a similar potential is shown
  to affect only quantitatively the temperature dependence of the
  conductance.  As expected, in the Kondo regime the numerical results
  are in excellent agreement with the mapped conductances. In the
  mixed-valence regime, the mapping describes accurately the
  low-temperature tail of the conductance.  The mapping is shown to
  provide a unified view of conduction in the single-electron
  transistor.
\end{abstract}
\pacs{73.23.-b,73.21.La,72.15.Qm,73.23.Hk}

\maketitle

\section{Introduction}
\label{sec:intro}
The development of the first single-electron transistor
(SET)\cite{GSM+98.156} was preceded by analytical and numerical
breakthroughs\cite{GR87:452,NL88.1768,HDW91:3720,MW92:2512,WM94:11040,CHZ94.19}
and, in the subsequent years, motivated numerous theoretical
investigations.\cite{ISS98:2444,HKS01:156803,AL03.245307,FFA03:155301,PuG04:R513,%
  ZB06.035332,NO06.125108,MNU04:3239,RWH+06:196601,SSI+06_096603,BCP08:395}
Notwithstanding this intense activity, important aspects of the
transport properties of the device received limited attention. In
particular, even after it guided the interpretation of experimental
data drawn out of
nanodevices,\cite{GGK+98.5225,CLGp02:226805,SAK+05:066801,KSA+06:36}
the concept of universality remained confined to the narrow corner in
which it was first identified.\cite{CHZ94.19} 

As it was established much more recently, in the Kondo regime of the Anderson
model,\cite{An61:41} the thermal dependence of the zero-bias
conductance of quantum dots either embedded\cite{SYO2009} in or
side-coupled\cite{SYO09:000} to the conduction path of elemental
nanodevices maps linearly onto the conductance curve for the symmetric
model computed by Costi, Hewson and Zlatic.\cite{CHZ94.19} The linear
coefficient in the mapping, which is parametrized by the ground-state
phase shift $\delta$ of the conduction electrons to which the quantum
dot is coupled, depends on the geometry. Qualitatively different
thermal dependences result, which distinguish the side-coupled
geometry from the embedded configuration. With a side-coupled quantum
dot, the linear coefficient depends only on the phase shift.
As it results, the application of a potential $\vw$ to the conduction
electrons can switch the conductance curves from monotonically
increasing to monotonically decreasing functions of the
temperature.\cite{SYO09:000}

In the embedded configuration, of which the SET is the simplest
illustration, the conductance always decreases with temperature. The
mapping depends on the difference $\delta-\deltaw$ between the
ground-state phase shift and the phase shift $\deltaw$ that the
potential $\vw$ would induce if the conduction electrons were
decoupled from the dot. In the special class of model Hamiltonians
that are invariant under particle-hole transformations, the symmetry
makes the dot occupation $n_d$ unitary, and it follows from the
Friedel sum rule\cite{La66:516} that $\delta=\pi/2$. For asymmetric
Hamiltonians, by contrast, a priori knowledge of $\delta$ is
impossible. In the Kondo regime, again on the basis of the Friedel sum
rule, a difference $\delta-\deltaw$ close to $\pi/2$ is
expected;\cite{SYO2009} accurate estimates nevertheless require
diagonalization of the model Hamiltonian. One possibility would be to
generate temperature-dependent conductance curves from
Bethe-ansatz\cite{AFL83:331,TW83:453} results for the phase
shift. Here, however, we prefer the numerical renormalization-group
(NRG) approach, which gives direct access to the phase shifts and to
the thermal dependence of the conductance.

This paper compares numerically computed SET conductances with the
mapping to the universal function. From the same NRG
diagonalization of the model Hamiltonian, we compute (i) the
conductance as a function of the temperature; and (ii) the
ground-state phase shift, which specifies the mapping.  Plotted as
functions of the temperature $T$ in the Kondo regime, the computed
conductances rise from from nearly zero to nearly ballistic as $T$
decreases past the Kondo temperature, and conduction electrons screen
the the dot magnetic moment. The curves obtained from the mapping run
through the numerical data.

Outside the Kondo domain, the mapping to the univeral function
is expected to fail. To witness its downfall, the numerical survey
includes the adjacent mixed-valence domain, in which the dot moment is
only partially formed. Here, the conductance crosses over to its
ground-state value at relatively high temperatures, which are outside
the domain of the mapping; hence, only the low-temperature tail of the
numerical data can be accurately fitted.

Our presentation is distributed over five
Sections. Section~\ref{sec:model} describes the SET and the Anderson
Hamiltonian modeling it. Section~\ref{sec:map} discusses cursorily the mapping
derived in Ref.~\onlinecite{SYO2009}. Section~\ref{sec:numeric} is
dedicated to the NRG procedure, which it summarizes, and to the
numerics, which it details. The numerical results appear next, in
Section~\ref{sec:result}, which starts with a discussion of the
phase-shift differences $\delta-\deltaw$.

The same section presents the conductance curves. Besides describing
quantitatively the conductance in the Kondo regime (and the
low-temperature sector of the mixed-valence regime), the mapping to
the universal function offers a simple, unifying view of charge transport
through a single-electron transistor. Faithful to this notion,
Sections~\ref{sec:w=0} and \ref{sec:w!=0} compound the output of 100
NRG runs in two plots of the conductance as a function of the
temperature and gate potential applied to the quantum dot; and show
that every feature of the landscapes is easily understood in the light
of that mapping. Section~\ref{sec:thermal-cond} then turns the
numerically computed ground-state phase shifts into linear
coefficients, and the mapping into conductance curves that fit the NRG
results for the temperature-dependent conductances. Our conclusions
and a summary constitute the closing Section~\ref{sec:conclude}.

\section{Model}
\label{sec:model}
\begin{figure}[th]
  \centering
  \includegraphics[width=\columnwidth]{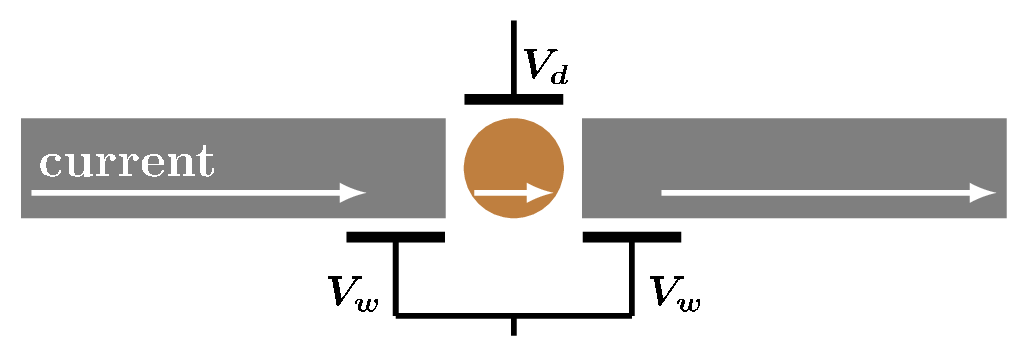}
  \caption[\ ]{(Color online) Single-electron transistor. The gate
    potentials $V_g$ and $V_w$ control the dot energy $\ed$ and the
    energy of the Wannier orbital $\fzero$, respectively. Symmetric
    tunneling amplitudes $V$ couple the dot to the quantum wires and
    allow conduction along the indicated path.}
  \label{fig:1}
\end{figure}

Figure~\ref{fig:1} depicts the object of our study, a quantum dot
symmetrically coupled to two otherwise independent quantum wires. The
tunneling amplitude $V$ transfer charge between the dot and the
wires. To represent the dot, we introduce a single, spin-degenerate
level $\cd$. An energy $\ed$, controlled by the gate potential $V_d$,
and a Coulomb repulsion $U$ define the dot Hamiltonian $H_d$, which
can be written\cite{KWW80:1044}
\begin{equation}
  \label{eq:hdot}
  H_d = (\ed + \frac{U}2)n_d -\frac{U}2(n_{d\uparrow}-n_{d\downarrow})^2,
\end{equation}
to emphasize that the energy $\ed+U/2$ breaks the particle-hole
symmetry of the dot Hamiltonian.

Inclusion of the quantum wires, represented by a structureless
half-filled band of width $2D$ containing $N$ conduction states, and
of their coupling to the quantum dot yields the Anderson Hamiltonian
\begin{equation}
  \label{eq:hand}
  \ha=\sum_{k}\epsilon_k a_{k}^\dagger a_{k} +\vw f_0^\dagger \fzero
  +V(f_0^\dagger \cd+\hc)+H_{d},
\end{equation}
where
\begin{equation}
  \label{eq:f0}
    \fzero\equiv\sum_k a_k/{\sqrt N},
\end{equation}
and $\vw$ is the potential due to the wire electrodes in
Fig.~\ref{fig:1}. 

The coupling $V$, to the wires, broadens the dot level $\cd$. The
scattering potential $\vw$ reduces its width $\Gamma=\pi\rho V^2$,
where $N\rho$ is the density of conduction states, to\cite{SYO2009}
\begin{equation}
  \label{eq:gammaW}
    \gammaw = \frac{\Gamma}{1+\pi^2\rho^2W^2}.
\end{equation}

With $\vw=0$, a specially important instance of Eq.~(\ref{eq:hand}) is
the particle-hole {\em symmetric Hamiltonian}
\begin{equation}
  \label{eq:hsym}
  \hs=\sum_{k}\epsilon_k a_{k}^\dagger a_{k} +\vw f_0^\dagger \fzero
  +V(f_0^\dagger \cd+\hc)-\frac{U}2(n_{\uparrow}-n_{\downarrow})^2,
\end{equation}
to which $\ha$ reduces for $\ed+U/2=W=0$.

The operators $a_k$ on the right-hand side of Eq.~(\ref{eq:hand}) are
even combinations of the conduction operators in the left and in the
right wire segments in Fig.~\ref{fig:1}:
$a_k=(c_{kL}+c_{kR})/\sqrt2$. The inversion symmetry of the device
decouples the odd combinations $(c_{kL}-c_{kR})/\sqrt2$ from the dot
level, which make no explicit contribution to the zero-bias
conductance $G(T)$. Linear Response relates $G(T)$ to the
dot-level spectral density $\rho_d$:\cite{SYO2009}
\begin{equation}
  \label{eq:glin}
    G(T) = \gc\,\pi\,\gammaw\,\int_{-D}^{D}\rho_{d}(\epsilon,T) \left[-\frac{\partial
    f(\epsilon)}{\partial\epsilon}\right]\,d\epsilon,
\end{equation}
where $\gc\equiv2e^2/h$ is the quantum conductance, $f(\epsilon)$ is
the Fermi function, $\za$ is the partition function for the
Hamiltonian $\ha$, and
\begin{equation}
  \label{eq:rhod}
  \rho_d(\epsilon,T)= \frac1{\za}\sum_{mn}\frac{e^{-\beta
      E_m}}{f(\epsilon)}|\matel{n}{\cdd}{m}|^2
  \delta(E_ m-E_n-\epsilon).
\end{equation}
Here, $\ket{m}$ ($\ket{n}$) denotes an eigenstate of $\ha$ with eigenvalue
$E_m$ ($E_n$).  

The substitution of Eq.~(\ref{eq:rhod}) on the right-hand side of
Eq.~(\ref{eq:glin}) yields an expression that translates into a
few lines of computer code:
\begin{equation}
  \label{eq:thermalG}
      G(T) = \gc\,\frac{\beta\pi\gammaw}{\za}\,
\sum_{mn}\frac{|\matel{m}{\cd}{n}|^2}{e^{\beta E_m}+e^{\beta E_n}}.
\end{equation}
As this expression suggests, the computational efforts underlying a
conductance curve $G(T)$ and \eg\ a magnetic susceptibility
plot\cite{KWW80:1003} are comparable.

\section{Mapping}
\label{sec:map}
Equation~(\ref{eq:thermalG}) yields conductances ranging from zero to
the quantum conductance $\gc$. We are interested in the Kondo regime,
the vast parametric subspace defined by the condition
$\gammaw\ll\min(|\ed|, U+\ed)$ and $k_BT
\ll\min(|\ed|, U+\ed, D)$. At the high-temperature end of the Kondo
regime a magnetic moment arises, associated with the nearly unitary
dot-level occupancy $n_d$. As the temperature is lowered past the
Kondo temperature $T_K$, and the wire electrons screen the resulting
dot magnetic moment, the Kondo cloud couples the dot level to the wire
states and sustains conductance. Costi, Hewson and
Zlatic\cite{CHZ94.19} showed that the thermal dependence of the
conductance for the symmetric Hamiltonian~(\ref{eq:hsym}) is a
universal function $\gset(T/T_K)$ of the temperature scaled by the
Kondo temperature. More recently, we have shown that the Kondo-regime
conductance maps linearly onto $\gset(T/T_K)$:\cite{SYO2009}
\begin{equation}\label{eq:guniversal}
  G\lp\frac{T}{T_K}\rp -\frac{\gc}2 = -
\lp\gset\lp\frac{T}{T_K}\rp-\frac{\gc}2\rp\cos2(\delta-\delta_W),
\end{equation}
where $\delta$ is the ground-state conduction-band phase shift, and
$\delta_w$, the Fermi-level phase shift for $V=0$, \ie\
\begin{equation}
  \label{eq:deltaw}
  \tan\delta_w= -\pi\rho W.
\end{equation}

As the temperature rises, the universal function decays from
$\gset(T\ll T_K)=\gc$, through $\gset(T=T_K)=\gc/2$, to $\gset(T\gg
T_K)=0$. The linear coefficient on the right-hand side of
Eq.~(\ref{eq:guniversal}) is a function of the phase-shift difference
$\delta-\deltaw$. It follows from the Friedel sum rule that
$2(\delta-\deltaw)=n_d\,\pi$, so that in the Kondo regime, and even in
the mixed-valence regime, $\delta-\deltaw$ is never small. According
to Eq.~(\ref{eq:guniversal}), for $\delta-\deltaw=\pi/2$, the
conductance $G(T/T_K)$ sticks to $\gset(T/T_K)$. A particular case is
the symmetric Hamiltonian~(\ref{eq:hsym}), for which $\delta=\pi/2$,
$\deltaw=0$, and $G(T/T_K)=\gset(T/T_K)$. If the difference were
$\delta-\deltaw=\pi/4$, on the other hand, the conductance would be
flat: $G(T)=\gc/2$. For the intermediate differences
$\pi/4<\delta-\deltaw\le \pi/2$ observed in the Kondo and
mixed-valence regimes, the conductance is a monotonically decreasing
function of the temperature. If $\delta-\deltaw\ne\pi/2$, $G(T\ll
T_K)$ is smaller than the quantum conductance, $G(T\gg T_K)$ is
nonzero, and $G(T/T_K)$ is flatter than $\gset(T/T_K)$.

Equation~(\ref{eq:guniversal}) offers a qualitative view of the
thermal dependence of the conductance, which combined with the Friedel
sum rule, describes $G(T/T_K)$ semiquantitatively.\cite{SYO2009} A
more attractive alternative, the numerical evaluation of $\delta$ and
$T_K$ is discussed next.

\section{Numerics}
\label{sec:numeric}
Excellent descriptions of the NRG method being
available,\cite{KWW80:1003,hewson93,BCP08:395} brief recapitulation of
the four steps constituting the procedure will be sufficient. Two
dimensionless parameters $\Lambda>1$ and $0<z\le 1$ define the {\em
  logarithmic discretization} of the conduction
band.\cite{YWO90:9403,BCP08:395} The infinite energy sequence
$\mathbb{E}_{m}=D\Lambda^{1-z-m}$ ($m=0,1,\ldots$) defines the
intervals $\mathbb{I}_{m}= [\mathbb{E}_{m+1}, \mathbb{E}_{m}]$. For
each interval, a single operator
$a_{m+}=\rho\int_{\mathbb{I}_{m}}\,a_k\,d\epsilon_k/n_{m}$, with
normalization factor $n_m$, is defined. In the negative half of the
conduction band, the sequence $-\mathbb{E}_{m}$ ($m=0,1\ldots$)
defines the mirror image $a_{m-}$ of each operator $a_{m+}$. The
$a_{m\pm}$ forma a basis upon which the conduction band Hamiltonian is
projected.\cite{KWW80:1003}

Next, a {\em Lanczos transformation}\cite{lanczos} makes tridiagonal
the projected conduction Hamiltonian, so that the model Hamiltonian
reads
\begin{equation}
  \label{eq:lanczos}
  \ha =\lp\sum_{n=0}^{\infty} t_n f_{n}^\dagger f_{n+1}
  + V f_0^\dagger \cd +\hc \rp  +\vw f_0^\dagger \fzero +H_{d}.
\end{equation}
Here, $\fzero$ is the operator defined in Eq.~(\ref{eq:f0}), and the
$f_n$'s ($n=0,1,\ldots$) form an orthonormal basis that replaces the
$a_{m\pm}$'s ($m=0,1,\ldots$). With $z=1$, we recover the Lanczos
transformation in Ref.~\onlinecite{KWW80:1003}. Otherwise, the
codiagonal coefficients $t_n$ have to be determined
numerically.\cite{YWO90:9403} With error $\mc{O}(\Lambda^{-n})$, it is
found\cite{CO05.104432} that $t_n=
D[(1-\Lambda^{-1})/\log\Lambda]\Lambda^{1-z-n/2}$. This shows that the
$t_n$'s decrease rapidly with $n$, a conclusion that brings us to the
third step in the NRG procedure, the {\em definition of a
  renormalization-group transformation}.

Given a temperature $T$ and a small dimensionless parameter $\alpha$,
let $N$ be the smallest integer such that $\ti_N<\alpha k_BT$, and
consider the infinite sum on the right-hand side of
Eq.~(\ref{eq:lanczos}). Compared to $k_BT$, the codiagonal element
$t_{N}$ is then negligible, and to compute $G(T)$, it is safe to
neglect the term with $n=N$. This decouples the subsequent terms in
the sum from the quantum dot, so that they no longer contribute to the
conductance. The infinite sum can therefore be truncated at
$n=N-1$. We define the reduced bandwidth $D_N\equiv
D[(1-\Lambda^{-1})/\log\Lambda]\Lambda^{-(N-1)/2}$ and the
dimensionless, scaled, truncated Hamiltonian $\ha^N$:
\begin{equation}
  \label{eq:hnrg}
 D_N\ha^N \equiv\lp\sum_{n=0}^{N-1} t_nf_{n}^\dagger
   f_{n+1} + V f_0^\dagger \cd+\hc\rp 
+\vw f_0^\dagger \fzero +H_{d}.
\end{equation}
The entire Hamiltonian has been divided by $D_N$ so that the
smallest codiagonal coefficient in the scaled sum, $t_{N-1}/ D_{N}$,
is of $\mathcal{O}(1)$.

The last step in the NRG procedure is the \emph{iterative
  diagonalization} of the model Hamiltonian. With $N=0$, the
right-hand side of Eq.~(\ref{eq:hnrg}) is easily diagonalized; four
eigenvalues $E_m^0$ and four eigenvectors $\ket{m}_{0}^{\phantom{A}}$
($m=1,\ldots,4$) result. At this stage, it is equally simple to
calculate the matrix elements
${}_0^{\phantom{A}}\!\matel{m}{\cd}{n}_0^{\phantom{A}}$ between the
eigenvectors of $H_A^{N=0}$, which will be needed to compute the
right-hand side of Eq.~(\ref{eq:thermalG}).

Application of the operators $f_{0\uparrow}^\dagger$,
$f_{0\downarrow}^\dagger$, $f_{0\uparrow}^\dagger
f_{0\downarrow}^\dagger$, and the identity $\openone$ on the
eigenvectors of $H_A^{N=0}$ generates sixteen states that constitute a
basis upon which the Hamiltonian $H_A^{N=1}$ can be
projected. Appropriately chosen linear combinations of those operators
yield basis states $\ket{p}_1$ ($p=1,\ldots,16$) that diagonalize the
charge and spin operators; projected on them, $H_A^{N=1}$ reduces to
block-diagonal matrices, which are then diagonalized numerically. The matrix
elements ${}_0^{\phantom{A}}\!\matel{m}{\cd}{n}_0^{\phantom{A}}$
($m,n=1,\ldots,4$) are projected onto the basis $\ket{p}_1$ and
subsequently rotated to the basis of the eigenstates $\ket{m}_{1}$
($m=1,\ldots,16$) of $H_A^{N=1}$. Application of the operators
$f_{1\uparrow}^\dagger$, $f_{1\downarrow}^\dagger$,
$f_{1\uparrow}^\dagger f_{1\downarrow}^\dagger$, and $\openone$ on the
$\ket{m}_{1}$ creates 64 basis vectors upon which $H_A^{N=2}$ can be
projected, and the procedure is iterated.

To check the exponential growth of matrix dimensions, a dimensionless
parameter $\upsilon$ is chosen, which will control the cost and the
accuracy of the iterative diagonalization. At the end of iteration
$N$, the eigenvectors with scaled energies $E_m/D_N$ above $\upsilon$
are discarded before the construction of the basis states
$\ket{p}_{N+1}$, upon which the Hamiltonian $\ha^{N+1}$ will be
projected. This expedient limits the number of basis states and hence
the computational effort that each iteration requires. The cost of a
full NRG run grows linearly with the number of iterations.

The diagonalization yields scaled eigenvalues ranging from unity to
$\upsilon$, \ie\ a window of energies ranging from $D_N$ to $\upsilon
D_N$. Having neglected $\ti_N\approx D_{N+1}$, we can only compute
conductances for $k_BT> \gamma D_{N+1}$, where $\gamma \agt 10$. At
the other extreme, the ultraviolet truncation restricts us to
temperatures such that $k_BT <\upsilon D_N$. Thus, provided that
$(D_N\upsilon)/(\gamma D_{N+1}) >\sqrt{\Lambda}$, \ie\ that
$\upsilon>\gamma$, the $N$-th iteration yields reliable conductances
in the temperature window $\gamma D_{N+1} \le k_BT \le
\sqrt{\Lambda}\gamma D_{N+1}$. If a run is stopped at iteration
$N_{max}$, the juxtaposition of the resulting windows yields $G(T)$
for all temperatures above $\gamma D_{N_{max}+1}/k_B$. In practice,
the conductance is only computed for the $k_BT\le0.1\,D$, because
irrelevant operators artificially introduced by the
logarithmic discretization make the interval $0.1\,D<k_BT \le D$
unreliable.

Conductance curves computed with large $\Lambda$ show oscillations,
which can be traced to a sequence of poles on the $\Im\epsilon=\pm
i\pi/\log{\Lambda}$ lines of the complex-energy
plane.\cite{YWO90:9403} To eliminate these artifacts of the
discretization, we average the conductance curve $G(T)$ computed for
given $z$ over a sequence of equally spaced $z$'s in the interval
$0<z\le 1$.\cite{OO94:11986} The exponential deependence of the
computational effort on $1/\log\Lambda$, makes this averaging
procedure far more efficient than comparably accurate computations
with small $\Lambda$.

The conductances in Section~\ref{sec:result} were computed with
$\Lambda=6$ and averaged over two $z$'s: 0.5 and 1. The amplitude of
the residual oscillations encountered after averaging over $z$,
somewhat smaller than $0.001\,e^2/h$, provides an estimate of the
error introduced by the logarithmic discretization. The other two
parameters controlling the precision of the results were fixed at
$\gamma=10.5$ and $\upsilon=50$, respectively. Spin degeneracies not
counted, the number of states below the cutoff in each iteration
peaked at 4000 in iteration 6. To estimate the error due to the
infrared and ultraviolet truncations, for each $N< N_{max}$ we
compared the conductance at the lowest temperature in the ($N-1$)-th
window, $\gamma D_{N} \le k_BT \le \sqrt{\Lambda}\gamma D_{N}$, with
the conductance at the highest temperature in the $N$-th window. The
mismatch between the two results never exceeding $0.001\,e^2/h$, we
conclude that deviations due to the three approximations in the
procedure, the logarithmic discretization and the infrared and
ultraviolet truncations, are comparable. At any temperature, the
estimated absolute deviation in the computed conductances is smaller
than than $0.05\%$ of the quantum conductance.

The relatively large discretization parameter expedites the
calculation. On a standard desktop computer, a complete run, including
(i) the iterative diagonalization of $\ha$ and computation of the
matrix elements on the right-hand side of Eq.~(\ref{eq:thermalG}) for
for each $z$, and (ii) the evaluation of the conductance curve in the interval
$10^{-10}\,D< k_BT \le 0.1\,D$, takes less than 30 seconds.

\subsection*{Phase shifts}
\label{sec:phase-shifts}

Along with the iterative diagonalization procedure,
Eq.~(\ref{eq:hnrg}) defines a renormalization-group transformation:\cite{KWW80:1003}
\begin{equation}\label{eq:tNRG}
\mathbb{T}[H_A^N] \equiv H_A^{N+2}=\Lambda H_A^{N}+
\sum_{n=N}^{N+1}\frac{t_n}{D_{N+2}} \lp f_{n}^\dagger f_{n+1} + \hc\rp.
\end{equation}
The factor $\Lambda$ multiplying the first term on the right-hand side
magnifies the scale on which the eigenvalues of $\ha^N$ are examined.
On the new scale, the second term is a fine structure. In the absence of
characteristic energies, as $N$ grows the magnification compensates
the refinement, and the lowest-energy eigenvalues of $\ha^{N+2}$
rapidly become indistinguishable from those of $\ha^{N}$. This
indicates that the Hamiltonian has reached a fixed point of
$\mathbb{T}$.

In the Kondo regime, the condition $V=0$ turns the Anderson
Hamiltonian $\ha$ into the local-moment fixed point (LM) of
$\mathbb{T}$. With $V\ne0$, as the temperature is reduced past the
dominant characteristic energy  $E_c=\min(|\ed|,\ed+U,D)$, the Hamiltonian
$H_{A}^N$ first approaches the LM and then moves away towards the
frozen-level fixed point (FL)---a strong-coupling fixed point
equivalent to Eq.~(\ref{eq:hand}) with $V\to\infty$. Between the LM
and the FL lies the Kondo temperature $T_K$, around which the
conduction electrons screen the dot moment.

If one of the dot excitation energies, $\Delta_0\equiv|\ed|$ or
$\Delta_2\equiv\ed+U$, is smaller than the dot width $\gammaw$, the
model Hamiltonian enters the mixed-valenc
regime.\footnote{Ref.~\onlinecite{KWW80:1044} discusses the
  mixed-valence regime under the heading ``transitional cases'', in
  Section III.E.}  Instead of $\min(D, |\ed|, |ed+U)$, the
dominant characteristic energy is now $E_c=\min(D, \gammaw)$. The dot moment is
only partially formed, as the coupling $\gammaw$ drives the model
Hamiltonian toward the FL before it can come close to the LM.

Devoid of characteristic energies, the two fixed points, LM and FL,
are phase-shifted conduction bands, which can be diagonalized
analytically.\cite{KWW80:1044} With $V=0$, for instance, the model
Hamiltonian flows towards the LM with phase shift $\deltaw$. With
$V\ne0$, (inaccurate) estimates for the LM phase shifts can be extracted
from the eigenvalues of $H_A^N$, where $N$ is such that $D_N\gg
k_BT_K$, \ie\ such that the model Hamiltonian dwells in the vicinity
of the LM.

Much more accurate FL phase shifts $\delta$ can be obtained from the
low-energy eigenvalues of $H_A^N$, because for large $N$ the
eigenvalues of $H_A^N$ come arbitrarily close to the many-body
excitations of the FL Hamiltonian\cite{KWW80:1044,SYO2009}
\begin{equation}
  \label{eq:hfl}
  H_{FL}^* =\sum_{\ell,\pm}\varepsilon_{\ell\pm}^*g_{\ell\pm}^\dagger g_{\ell\pm}.
\end{equation}
Here, the $+$ and $-$ subscripts distinguish the positive eigenvalues
from the negative ones, while $\ell=0,1,\ldots$ counts the positive
(negative) eigenvalues upward (downward) from the Fermi level. 

Once the eigenvalues of $H_A^N$ are identified with the many-body
energies generated from Eq.~(\ref{eq:hfl}), the ground-state phase
shift $\delta$ are extracted from the approximate expression
describing all but the $\varepsilon_{\ell\pm}^*$ closest to
zero.\cite{KWW80:1044} For $\Lambda \ge5$, in particular, within
$0.1\%$ deviation,
\begin{equation}
  \label{eq:etas}
  \varepsilon_{\ell\pm}^*=
  \pm\Lambda^{\nu+\ell\mp\delta/\pi}\qquad(\ell=1,2,\ldots),
\end{equation}
where $\nu=1-z$ ($\nu=3/2-z$) for odd (even) $N$. 

\section{Results}
\label{sec:result}

To emulate the conditions under which a SET operates, we fix the
Coulomb repulsion ($U=5\,D$) and effective dot-level width
($\gammaw=0.15\,D$) and examine the ground-state phase shift and the
thermal dependence of the conductance as a function of the dot energy
$\ed$ for five wire potentials $\vw$. Since $\ha$, $G$, and $|\delta|$
are invariant under the transformation $c_d\to -c_d^\dagger$, $a_k\to
a_k^\dagger$, $\ed+U/2\to -(\ed+U/2)$, $W\to-W$, we need not study
negative wire potentials, which would mirror the conductances and
phase shifts calculated with positive $W$.

\begin{figure}[h]
  \centering
  \includegraphics[width=\columnwidth]{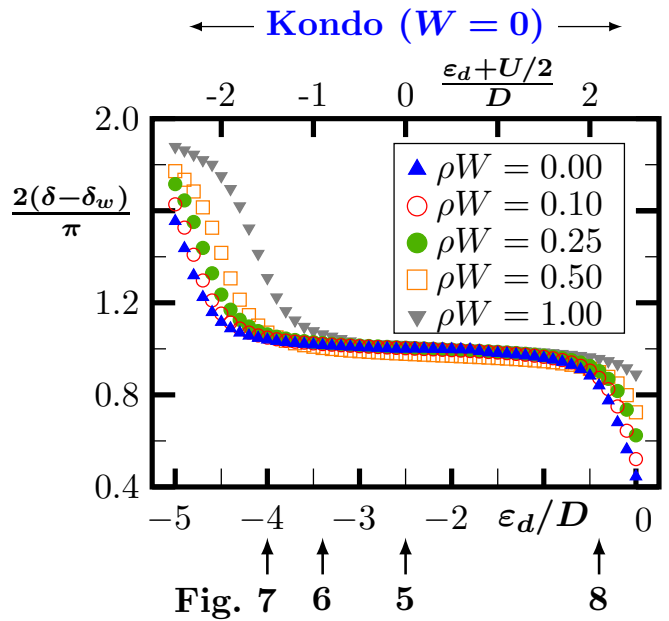}
  \caption{(Color online) Ground-state phase shift $\delta$, measured
    from the phase shift $\deltaw$ obtained from Eq.~(\ref{eq:deltaw})
    for the displayed wire potentials $\vw$, as a function of the
    dot-level energy $\ed$. The Friedel sum rule shows that the
    ordinate is equal to the dot occupation $n_d$. The $\delta$'s were
    obtained, with the help of Eq.~(\ref{eq:etas}), from the
    low-energy spectrum of $\ha$ resulting from NRG runs with $U=5\,D$
    and $\gammaw=0.15\,D$. The arrows above the top horizontal axis
    define the Kondo domain for $W=0$. For $W\ne0$, the Kondo domain
    is displaced to the right. Each vertical arrow pointing to the
    lower horizontal axis identifies the figure displaying the thermal dependence
    of the conductance for the indicated dot energy.}
  \label{fig:2}
\end{figure}

We turn first to the calculated phase shifts.  Figure~\ref{fig:2} displays the
argument of the trigonometric function on the right-hand side of the
mapping~(\ref{eq:guniversal}), computed for five wire potentials
$\vw$, in the dot-energy range $0\le \ed/D \le-U$. For
$\vw=0$, the upright triangles draw a well-known curve,\cite{TW83:453}
which remains invariant under the particle-hole transformation
$\ed+U/2\to -(\ed+U/2)$, $\delta\to\pi-\delta$. At the symmetric
point $\ed+U/2=0$, which corresponds to Eq.~(\ref{eq:hsym}), the
phase-shift is exactly $\pi/2$. The arrows above the top axis indicate
the Kondo domain, within which the phase shift remains close to
$\pi/2$. As $|\ed+U/2|$ grows, the model Hamiltonian first approaches
the limits of the Kondo domain and then invades the mixed-valence
domain. In response, $\delta$ moves away from $\pi/2$, towards zero
for $\ed+U/2\to U/2$, or towards $\pi$ for $\ed+U/2\to-U/2$.

The wire potential reduces the ground-state phase shift throughout the
depicted range. For $\rho\vw=1$, for instance, the phase shift at the
symmetric dot-level energy $\ed=-2.5\,D=-U/2$ is reduced from
$\delta=\pi/2$ to $\delta=\pi/10$. In the Kondo-regime, as the
illustration shows, the difference $\delta-\deltaw$ is nonetheless
pinned at $\pi/2$.  The pinning is due to the Friedel sum
rule.\cite{La66:516} Since the ground-state phase shift would be
$\delta_w$ if $V$ were zero, $2(\delta-\delta_w)/\pi$ is the screening
charge due to the coupling to the dot. In the Kondo regime, that
charge is nearly unitary, and $\delta-\delta_w\approx\pi/2$.

Figure~\ref{fig:2} shows that a positive wire potential tends to
displace the Kondo regime towards higher dot energies. For $\vw=0$,
the rapid decay of the phase shift near $\ed=0$ ($\ed=-U$) marks the
resonance between the $n_d=0$ and $n_d=1$ ($n_d=1$ and $n_d=2$)
dot-level configurations. The Kondo regime lies between them. As $\vw$
grows, the two resonances move to higher $\ed$'s, and so does the
Kondo regime.

\subsection{Conductance landscape}
\label{sec:w=0}

\begin{figure}[th]
  \centering
  \includegraphics[width=\columnwidth]{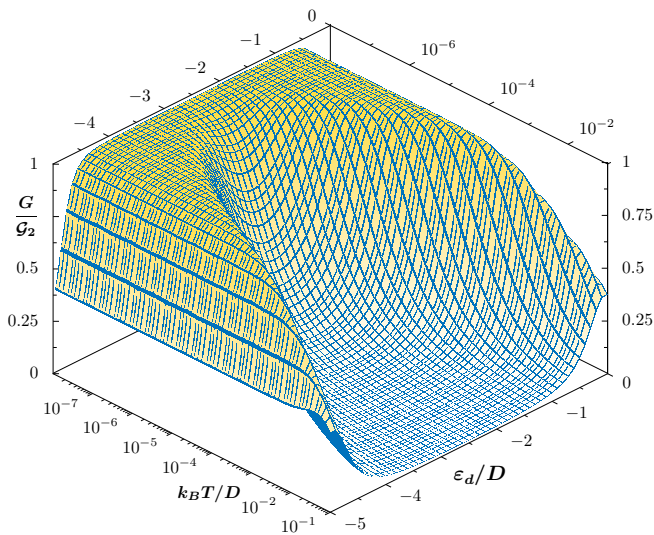}
  \caption{(Color online) Conductance as a function of the temperature
    and dot-level energy, for $U=5\,D$, $\Gamma=0.15\,D$, and
    $W=0$. The plot is symmetric with respect to the
    $\ed=-U/2=-2.5\,D$ plane. The sharp drops near $\ed=-5\,D$ and
    $\ed=0$ mark the borders of the Kondo regime, which extends
    roughly from $\ed=-\Gamma$ to $\ed+U=-\Gamma$. In the Kondo
    regime, at fixed $\ed$, the more gradual decay of the conductance
    with the temperature portrays the evaporation of the Kondo droplet.}
  \label{fig:3}
\end{figure}

According to Eq.~(\ref{eq:guniversal}), $\delta-\deltaw$ controls
$G(T)$. Consequently, the central features of Fig.~\ref{fig:2} are
manifest in landscape plots of the conductance. Figure~\ref{fig:3}
shows $G(T)$ in the dot-energy range $|\ed+U/2| \le U/2$ for $U=5\,D$,
and $\Gamma=0.15\,D$. The plot surveys the entire Kondo regime and
part of the mixed-valence regime. The plane $\ed=-U/2$, which represents
the symmetric Hamiltonian~(\ref{eq:hsym}), splits the landspace in two
symmetric halves, mapped onto each other by the particle-hole
transformation $c_d\to-c_d^\dagger$, $a_k\to a_k^\dagger$.

Examined at the symmetric point $\ed=-U/2$, the temperature-dependent
conductance reproduces the universal function $\gset(T/T_K)$. Here and
elsewhere in the Kondo regime, the conductance at fixed $\ed$ rises
from zero to ballistic or nearly ballistic as the temperature is
reduced past $T_K$, \ie\ as one climbs from the high-temperature
Coulomb-blockade valley to the low-temperature Kondo plateau. The
Kondo temperature for the symmetric Hamiltonian is
$8\times10^{-7}\,D$. As $|\ed+U/2|$ grows, it rises until
$k_BT_k\approx 0.1\,D$, an equality indicating proximity to the
mixed-valence regime, \ie\ to the two resonances centereced at
$\ed=-5\,D$ and $\ed=0$. As $|\ed+U/2|$ grows further, we come into
mixed-valence domain. The dot moment shrinks, and so does the Kondo
cloud. The Kondo bypass of the Coulomb blockade becomes less and less
effective, and the conductance approaches zero. The steep drops near
the $\ed=0$ and $\ed=-5$ planes in Fig.~\ref{fig:3} mark the
mixed-valence regime.

\subsection{Wire potential}
\label{sec:w!=0}

\begin{figure}[th]
  \centering
  \includegraphics[width=\columnwidth]{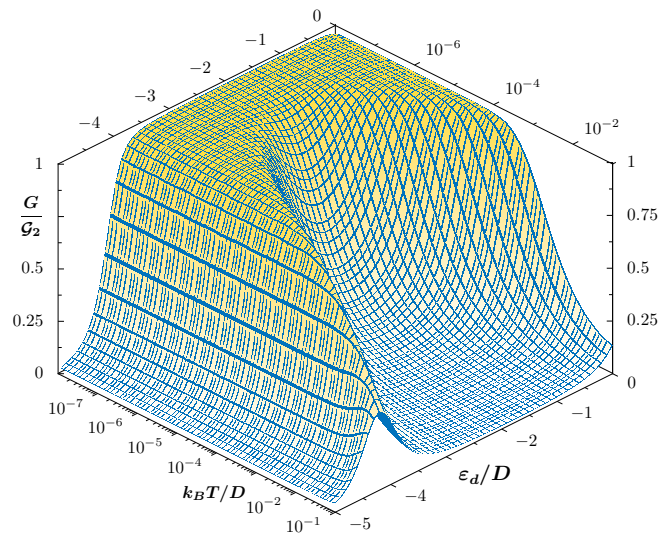}
  \caption{(Color online) Conductance as a function of the temperature
    and dot-level energy for $U=5\,D$, $\gammaw=0.15\,D$, and $\rho
    W=0.50$. The wire potential breaks the particle-hole symmetry
    visible in Fig.~\ref{fig:3}. The sharp drop centered at
    $\ed=-5\,D$ in Fig.~\ref{fig:3} is now fully visible, while the
    one centered at $\ed=0$ is out of sight, an indication that the
    Kondo regime has been displaced to higher dot energies. The
    bell-shaped resonance near the bottom left corner of the
    $k_BT=10^{-1}\,D$ plane stakes the mixed-valence regime.}
  \label{fig:4}
\end{figure}

Figure~\ref{fig:4} displays the conductance as a function of $\ed$ and
$T$ for $U=5\,D$, $\gammaw=0.15\,D$, and $\rho W=1$. Quantitative
differences distinguish the plot from Fig.~\ref{fig:3}. In particular,
the Kondo temperature is now minimized at the higher dot-level energy
$\ed=-1.9\,D$, the minimum $k_BT_K=2.4\times10^{-6}\,D$ is thirtyfold
higher, the resonance between the $n_d=1$ and $n_d=2$ dot
configurations is now centered at $\ed\approx-4.2\,D$, and the
resonance between the $n_d=0$ and $n_d=1$ dot configurations has been
reduced to an incipient rise, at the high-$\ed$ end of the
plot. Clearly, the wire potential has displaced the Kondo domain
towards higher dot-level energies. This displacement acknowledged, we
recognize in Fig.~\ref{fig:4} the salient features of
Fig.~\ref{fig:3}.

The two landscapes are similar because the phase-shift difference on
the right-hand side of Eq.~(\ref{eq:guniversal}) is weakly dependent
on $\rho W$. With $\delta-\deltaw\approx\pi/2$, the conductance curve
$G(T/T_K)$ is approximately mapped onto $\gset(T/T_K)$ throughout the
Kondo domain. The rise from the high-temperature valley to the Kondo
plateau is therefore close to universal, dependent on the model
parameters only through the Kondo temperature $T_K$. The latter
is a function of the antiferromagnetic interaction $J$ between the dot
moment and the conduction electrons around it. The Schrieffer-Wolff
transformation\cite{SW66:491} relates that interaction to the dot
excitation energies:
\begin{equation}\label{eq:jsw}
\rho J = \frac{2\gammaw}{\pi}\lp\frac1{\Delta_0}+\frac1{\Delta_2}\rp.
\end{equation}
Here, $\Delta_0$ ($\Delta_2$) is the energy needed to remove (add) and
electron to the singly-occupied dot level. For the symmetric
Hamiltonian~(\ref{eq:hsym}), in particular,
$\Delta_0=\Delta_2=U/2$. For nearly symmetric Hamiltonians,
$\Delta_0=|\ed|$ and $\Delta_2=U+\ed$. As $|W|$ or $|\ed+U/2|$ grow,
the resulting particle-hole asymmetry renormalizes the dot
energy,\cite{Hal78:416,KWW80:1044} so that $\Delta_0$ and $\Delta_2$
are changed to $\Delta_0^*=|\ed^*|$ and $\Delta_2^*=U+\ed^*$,
respectively, where $\ed^*$ is the effective dot energy at the
LM.\cite{KWW80:1044}

Since both landscapes were computed for the same effective width
$\gammaw=0.15\,D$, only (i) the excitation energies $\Delta_0^*$ and
$\Delta_2^*$; and (ii) irrelevant operators make the Kondo temperatures
in Fig.~\ref{fig:3} different from those in Fig.~\ref{fig:4}. The
renormalized excitation energies displace the Kondo domain along the
$\ed$ axis, while the modified irrelevant operators extend the Kondo
plateau towards higher temperatures.

This concludes our overview of the numerically computed
conductances. Section~\ref{sec:thermal-cond} will inspect in more
detail the data in four slices of Figs.~\ref{fig:3}~and \ref{fig:4}
and compare them to Eq.~(\ref{eq:guniversal}).

\subsection{Thermal dependence of the conductance}
\label{sec:thermal-cond}
\begin{figure}[th]
  \centering
  \includegraphics[width=\columnwidth]{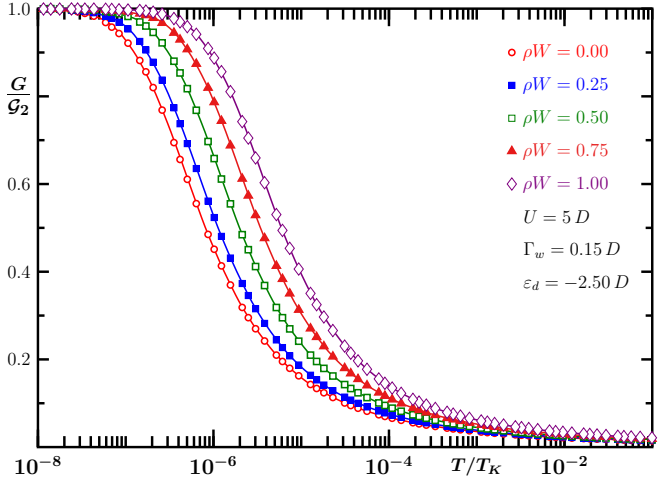}
  \caption{(Color online) Thermal dependence of the conductance for
    $\ed+U/2=0$, and the indicated values of the other model
    parameters. The circles, open and filled squares, triangles, and
    diamonds are NRG data, while the solid lines through them depict
    Eq.~(\ref{eq:guniversal}), with the Kondo temperatures and phase
    shifts listed in Table~\ref{tab:1}. The curve through the open
    circles, in particular, is the universal conductance
    $\gset(T/T_K)$ for the symmetric
    Hamiltonian~(\ref{eq:hsym}).\cite{CHZ94.19} Since
    $\delta-\deltaw\approx \pi/2$, each solid line is close to
    $\gset(T/T_K)$.}
\label{fig:5}
\end{figure}

Figure~\ref{fig:5} displays the conductance as a function of the
temperature for $U=5\,D$, $\gammaw=0.15\,D$, $\ed+U/2=0$, and five
wire potentials: $\rho W=0$ and $1$, already studied in
Figs.~\ref{fig:3}~and \ref{fig:4}, and three intermediate values,
$\rho W=0.25$, 0.5, and 0.75. With $W=0$, the open circles represent
the symmetric Hamiltonian~(\ref{eq:hsym}), and the solid line through
them reproduces the universal function $\gset(T/T_K)$ computed in
Ref.~\onlinecite{CHZ94.19}. Notwithstanding the wire potentials, the
Hamiltonians represented by the squares, triangles, and diamonds lie
deep inside the Kondo regime. For each of them, the phase-shift
difference $\delta-\deltaw$ in Table~\ref{tab:1} is close to
$\pi/2$. It follows that the right-hand side of
Eq.~(\ref{eq:guniversal}) is close to $\gset(T/T_K)$. The agreement
with the numerical data is excellent.

\begin{table}[th]
  \centering
  \begin{tabular}{ccrrrcl}
\hline\hline
Figure & Symbol & $\rho W$ & $\deltaw/\pi$ & $\delta/\pi$ 
& $\,2(\delta-\deltaw)/\pi$ &$k_BT_K/D$\\[3pt]
\hline
5&$\circ$&0.00 & 0.00 & 0.50 &  1.00 &  8.1\te{7}\\
5&$\blacksquare$&0.25 &-0.21 &  0.29 &  1.00 &  1.1\te{6}\\
5&$\square$&0.50 &-0.32 &  0.18 &  1.01 &    2.0\te{6}\\
5&$\blacktriangle$&0.75 &-0.37 &  0.13 &  1.01 &  3.4\te{6}\\
5&$\lozenge$&1.00 &-0.40 &  0.11 &  1.02 &    6.0\te{6}\\
6&$\circ$&0.00 & 0.00 & 0.51 &  1.02 &  4.4\te{6}\\
6&$\blacksquare$&0.25 &-0.21 &  0.30 &  1.03 &  1.1\te{5}\\
6&$\square$&0.50 &-0.32 &  0.20 &  1.03 &  3.6\te{5}\\
6&$\blacktriangle$&0.75 &-0.37 &  0.15 &  1.05 &  1.1\te{4}\\
6&$\lozenge$&1.00 &-0.40 &  0.13 &  1.06 &  3.6\te{4}\\
7&$\circ$&0.00 & 0.00 & 0.52 &  1.04 &  8.8\te{5}\\
7&$\blacksquare$&0.25 &-0.21 &  0.32 &  1.06 &  3.3\te{4}\\
7&$\square$&0.50 &-0.32 &  0.23 &  1.10 &  1.6\te{3}\\
7&$\blacktriangle$&0.75 &-0.37 &  0.22 &  1.17 &  7.3\te{3}\\
7&$\lozenge$&1.00 &-0.40 &  0.25 &  1.31 &  3.4\te{2}\,*\\
8&$\circ$&0.00 & 0.00 &  0.42 &  0.84 &  7.7\te{3}\\
8&$\blacksquare$&0.25 &-0.21 &  0.24 &  0.90 &  2.4\te{3}\\
8&$\square$&0.50 &-0.32 &  0.15 &  0.93 &  9.4\te{4}\\
8&$\blacktriangle$&0.75 &-0.37 &  0.10 &  0.95 &  4.0\te{4}\\
8&$\lozenge$&1.00 &-0.40 &  0.08 &  0.96 &  1.9\te{4}\\
\hline\hline
  \end{tabular}
  \caption{Phase shifts and Kondo temperatures for the twenty NRG runs
    depicted in Figs.~\ref{fig:5}-\ref{fig:8}. The ground-state phase
    shifts $\delta$ were obtained from Eq.~(\ref{eq:etas}), the wire
    phase shifts $\deltaw$, from Eq.~(\ref{eq:deltaw}), and the Kondo
    temperatures, from the definition $G(T= T_K)\equiv\gc/2$. As
    explained in the text, a different procedure identified the
    Kondo temperature marked with an asterisk, which is associated
    with a Hamiltonian in the mixed-valence regime.}
  \label{tab:1}
\end{table}

\begin{figure}[th]
  \centering
  \includegraphics[width=\columnwidth]{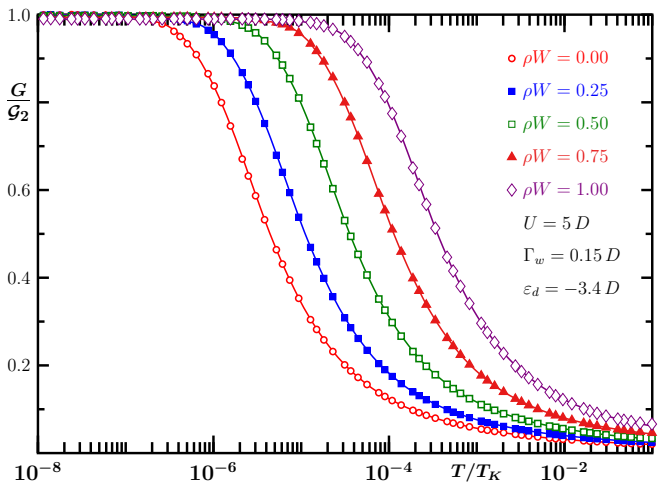}
  \caption{(Color online) Thermal dependence of the conductance for
    $\ed+U/2=-0.9\,D$. The symbols and lines were computed as
    described by the caption of Fig.~\ref{fig:5}. As Table~\ref{tab:1}
    shows, the argument $\delta-\deltaw$ on the right-hand side of
    Eq.~(\ref{eq:guniversal}) is close to $\pi/2$. As a consequence,
    the solid lines are only slightly different from
    $\gset(T/T_K)$. The agreement with the numerical data is, again,
    excellent.}
\label{fig:6}
\end{figure}

As the Hamiltonian moves away from the $\ed=-U/2$ plane, the
particle-hole asymmetry becomes more pronounced. One might expect the
difference $\delta-\deltaw$ to grow. As Fig.~\ref{fig:2} showed,
however, in the Kondo regime the Friedel sum rule restrains the
growth, so that $\delta-\deltaw\approx \pi/2$. Illustrative results
appear in Fig.~\ref{fig:6}, which displays conductance curves for
$\ed=-3.4\,D$. Even for the strongest wire potential in the legend,
$\rho W=1$, the difference $\delta-\deltaw$ in Table~\ref{tab:1} is
only 6\% away from $\pi/2$. As in Fig.~\ref{fig:5}, therefore, the
conductance curves computed from Eq.~(\ref{eq:guniversal}) are close
to $\gset(T/T_K)$. The agreement with the numerical data is again
flawless. Since we are now closer to the boundary of the Kondo regime,
the Kondo temperature is more sensitive to the renormalization of the
dot-level energy induced by strong wire potentials. Compared to
Fig.~\ref{fig:5}, Fig.~\ref{fig:6} thus exhibits a substantially
broader spread of crossover temperatures.

\begin{figure}[th]
  \centering
  \includegraphics[width=\columnwidth]{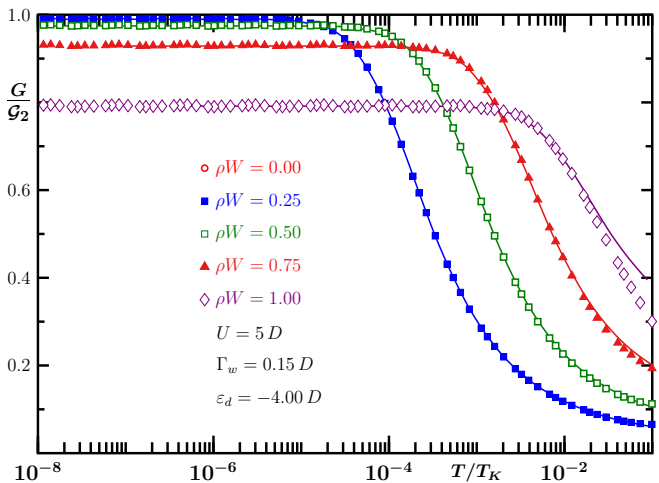}
  \caption{(Color online) Thermal dependence of the conductance for
    $\ed+U/2=-1.5\,D$. The symbols and lines were calculated as
    described by the caption of Fig.~\ref{fig:5}. As discussed in the text, the
    high-temperature separation between the solid line and the
    diamonds fingers a Hamiltonian outside the Kondo regime.}
\label{fig:7}
\end{figure}

Figure~\ref{fig:7} displays numerical results for $\ed+U/2=-1.5\,D$, a
still larger departure from the symmetric condition. For $\rho W\le
0.5$, the agreement with Eq.~(\ref{eq:guniversal}) is excellent; for
$\rho W=0.75$, it is imperfect only at the highest temperatures
shown. For $\rho W=1$, however, there is substantial disagreement, which
justifies a digression.

Inspection of Fig.~\ref{fig:4} shows that, for $\ed=-4\,D$ (and $\rho
W=1$), the model Hamiltonian lies well within the mixed-valence
regime.\footnote{With $\ed=-4\,D$ and $\rho W=0.75$, although in the
  mixed-valence regime, the model Hamiltonian is close to the Kondo
  regime.} In the Kondo regime, Eq.~(\ref{eq:guniversal}) is reliable
for thermal energies that are small on the scale of the dominant
characteristic energy $E_C=\min(|\ed^*|, U+\ed^*, D)$. If $\ed^*$ had
its bare value, $\ed=-4\,D$, the mapping would be reliable for
$k_BT\ll E_c=D$. The dot energy has been renormalized, however, and
the renormalization has pushed the model Hamiltonian into the
mixed-valence regime. The dominant characteristic energy has therefore
been changed to $E_c=\min(\gammaw, D)=\gammaw$, a reduction that
restricts the domain of the mapping to $k_BT \ll 0.15\,D$. The mapping
fails at higher temperatures because irrelevant operators, which are
sizable near the characteristic energy, make a significant
contribution $\Delta G_{\text{irr}}$ to the conductance. At
$k_BT=0.1D=2/3\gammaw$, for example, the diamonds in Fig.~\ref{fig:7}
are displaced $0.2e^2/h$ below the solid line; upon cooling, $\Delta
G_{\text{irr}}$ decays in proportion to $k_BT$ and becomes
insignificant below $k_BT=10^{-2}\,D$.

If $-\ed$ were steadily increased beyond $-\ed=4\,D$, the model Hamiltonian
would traverse the mixed-valence region. Once $|\ed^*+U|>\gammaw$, the
dot occupation would approach $n_d=2$. The dominant characteristic
energy $E_c=|\ed^*+U|$ would then define the crossover energy scale, which
would hence rise with $-\ed$. Soon, the model Hamiltonian would be
driven to the frozen-level fixed point at the first steps of the
renormalization-group flow, and the mapping would be reduced to its FL
limit, $G(T\to0)=\sin^2(\delta-\deltaw)\approx 0$.

\begin{figure}[th]
  \centering
  \includegraphics[width=\columnwidth]{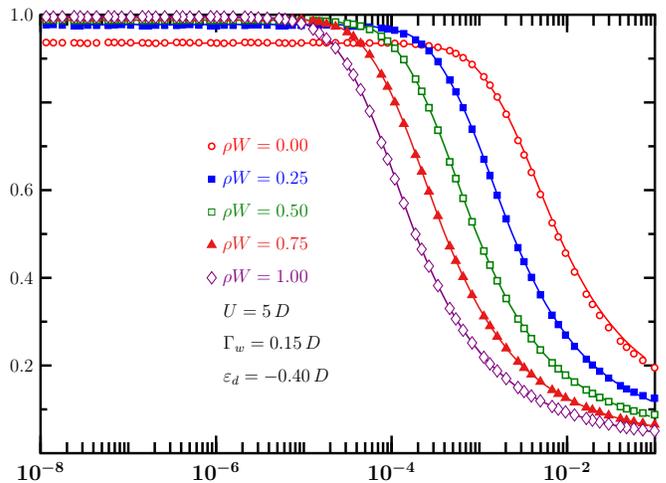}
  \caption{(Color online) Thermal dependence of the conductance for
    $\ed+U/2=2.1\,D$. The lines and symbols were computed as described
    by the caption of Fig.~\ref{fig:5}. The relatively large
    separation from the symmetric condition $\ed+U/2=0$ places the
    $W=0$ Hamiltonian close to the border of the Kondo regime; at high
    temperatures, relatively large irrelevant operators, whose
    influence decays in proportion to $k_BT/D$, introduce deviations
    from Eq.~(\ref{eq:guniversal}). Since the wire potential displaces
    the Kondo regime to higher dot-level energies, the distance to the
    border grows with $\rho W$. As a result, the squares, triangles, and diamonds
    are in excellent agreement with Eq.~(\ref{eq:guniversal}).}
\label{fig:8}
\end{figure}

Although Eq.~(\ref{eq:guniversal}) is asymptotically exact at low
temperatures, \ie\ for $k_BT\ll E_c$, as our digression showed its
practical value is eroded outside the Kondo regime. In the
mixed-valence regime, in particular, the asymptotic region lies below
the crossover temperature, \ie\ in the vicinity of the FL. To plot the
rightmost solid line in Fig.~\ref{fig:7}, we thus had to match the
right-hand side of Eq.~(\ref{eq:guniversal}) to the diamond at
$G=0.7\gc$, because the identification $G(T=T_K)=0.5\gc$, which
defined $T_K$ for all the other plots in
Figs.~\ref{fig:5}-\ref{fig:8}, became unreliable for $\rho W=1$. The
asterisk in Table~\ref{tab:1} marks the resulting Kondo temperature.

Near the opposite extreme of the Kondo regime, for fixed, small
$-\ed$, the wire potential drives the model Hamiltonian toward the
center of the Kondo regime. In Fig.~\ref{fig:4}, for instance, the
mixed-valence domain is missing, because the wire potential displaced
it to positive dot-level energies. Figure~\ref{fig:8} displays the
$\ed=-0.4\,D$ plane for the five potentials $\rho W=0$, 0.25, 0.5,
0.75, and 1. The $\rho W=0$ Hamiltonian is now at the boundary of the
Kondo regime, and for $k_BT>10^{-2}D$, irrelevant operators introduce
significant deviations $\Delta G_{\text{irr}}$ from the solid line. As
$\rho W$ grows, however, the model Hamiltonian sinks deeper into the
Kondo regime, and the agreement with the solid lines is recovered.

\section{Conclusions}
\label{sec:conclude}
In the Kondo regime, the physical properties for the Anderson model
are universal functions of the temperature scaled by the Kondo
temperature. The conductance constitutes no exception. For
$\ed+U/2=W=0$, the Anderson Hamiltonian reduces to
Eq.~(\ref{eq:hsym}), in which case the thermal dependence of the SET
conductance is the universal function
$\gset(T/T_K)$.\cite{CHZ94.19,BCP08:395} For nonzero $\ed+U/2$ or $W$,
Eq.~(\ref{eq:guniversal}) maps linearly the conductance onto
$\gset(T/T_K)$.\cite{SYO2009} With $\ed+U/2=W=0$, the particle-hole
symmetry of the Hamiltonian~(\ref{eq:hsym}) forces the ground-state
phase shift $\delta$ to be $\pi/2$ and reduces
Eq.~(\ref{eq:guniversal}) to a trivial identity. For asymmetric
Hamiltonians in the Kondo regime, as the 5th and 6th columns in
Table~\ref{tab:1} and the plots in Fig.~\ref{fig:2} suggest, while
the ground-state phase shift can take any value in its domain of
definition, the Friedel sum rule keeps the difference
$\delta-\deltaw$ close to $\pi/2$. The linear coefficient of the
mapping (\ref{eq:guniversal}) is never far from $-1$, and the thermal
dependence of the SET conductance, never far from $\gset(T/T_K)$.

Equation~(\ref{eq:guniversal}) becomes asymptotically exact as
$k_BT\ll E_c$. In the Kondo regime, the dominant characteristic energy
is $E_c=\min(D, |\ed^*|, U+\ed^*)$, and the mapping to the conductance
is reliable throughout the crossover from the LM to the FL. In the
mixed-valence regime, with $|\ed^*|<\gammaw$ or $U+\ed^*<\gammaw$, the
dominant characteristic energy $E_c=\min(D, \gammaw)$ reduces the
domain of the universal mapping to a temperature range close to the
low-temperature fixed point, \ie\ to the final steps in the rise to
the limit $G(T=0)=\gc\sin^2(\delta-\deltaw)$.\cite{SYO09:000}

\subsection*{Summary}
\label{sec:sum}

To recapitulate, the essentially exact numerical data in
Figs.~\ref{fig:2}-\ref{fig:8} offer an overview of electrical
conduction through a quantum dot embedded in the conductance path of a
nanostructured device. The linear mapping~(\ref{eq:guniversal}) to the
universal function $\gset(T/T_K)$ for the symmetric Anderson
Hamiltonian\cite{CHZ94.19,BCP08:395} describes accurately the thermal
dependence of the conductance for $k_BT\ll D$ in the Kondo regime. A
gate potential applied to the wires affects only quantitatively the
dependence of the conductance on the temperature and dot-level energy.

In the Kondo regime, independently of the wire potential, the Friedel
sum rule drives the thermal dependence of the conductance to the
neighborhood of $\gset(T/T_K)$. In particular, $G(T\ll
T_K)\approx\gc$. In the mixed-valence regime, only well below the
crossover temperature does $G(T/T_K)$ map onto $\gset(T/T_K)$, and
$G(T\to0)$ is substantially smaller than the quantum conductance.
In the laboratory, it is often difficult to distinguish the Kondo
regime from the mixed-valence regime. \cite{GGK+98.5225,SAK+05:066801}
The mapping to the universal curve offers a practical solution to this
problem.

\acknowledgments
This work was supported by the CNPq and FAPESP.


\end{document}